%% file: main.tex
\documentclass[sigconf,hyphens]{acmart}
\AtBeginDocument{%c``
  \providecommand\BibTeX{{%
    \normalfont B\kern-0.5em{\scshape i\kern-0.25em b}\kern-0.8em\TeX}}}

\usepackage{csquotes}
\usepackage{multirow}   
\usepackage{xcolor}
\usepackage{hyperref}
\usepackage[hyphenbreaks]{breakurl}

\usepackage[normalem]{ulem} % for strikethroughs, normalem 

\definecolor{editcolor}{RGB}{255,0,0} % Red for visibility, but choose any color you prefer

\copyrightyear{2024}
\acmYear{2024}
\setcopyright{rightsretained}
\acmConference[CHI EA '24]{Extended Abstracts of the 2024 CHI Conference on Human Factors in Computing Systems}{April 23--28, 2023}{Hamburg, Germany}
\acmBooktitle{Extended Abstracts of the 2024 CHI Conference on Human Factors in Computing Systems (CHI EA '23), April 23--28, 2023, Hamburg, Germany}\acmDOI{10.1145/3544549.3585697}
\acmISBN{978-1-4503-9422-2/23/04}

\begin{document}

\newcommand{\showURL}[1]{\texttt{#1}}
\title{A Value-Oriented Investigation of Photoshop’s Generative Fill}  

%https://news.microsoft.com/source/features/innovation/an-easier-way-to-help-people-give-tech-support-to-loved-ones-wins-microsoft-hackathon-top-prize/


\author{Ian P. Swift}
\affiliation{%
  \institution{University of Illinois Chicago}
  % \streetaddress{851 S. Morgan}
  % \city{Chicago}
   \state{Illinois}
   \country{USA}
  % \postcode{60607}
}
\email{iswift@uic.edu}

\author{Debaleena Chattopadhyay}
\affiliation{%
  \institution{University of Illinois Chicago}
  % \streetaddress{851 S. Morgan}
  \city{Chicago}
   \state{Illinois}
  \country{USA}
  % \postcode{60607}
}
\email{debchatt@uic.edu}

%\renewcommand{\shortauthors}{Chattopadhyay}

\begin{abstract}

The creative industry is both concerned and enthusiastic about how generative AI will reshape creativity. How might these tools interact with the workflow values of creative artists? In this paper, we adopt a value-sensitive design framework to examine how generative AI, particularly Photoshop’s Generative Fill (GF), helps or hinders creative professionals' values. We obtained 566 unique posts about GF from online forums for creative professionals who use Photoshop in their current work practices. We conducted reflexive thematic analysis focusing on usefulness, ease of use, and user values. Users found GF useful in doing touch-ups, expanding images, and generating composite images. GF helped users' values of productivity by making work efficient but created a value tension around creativity: it helped reduce barriers to creativity but hindered distinguishing `human' from algorithmic art. Furthermore, GF hindered lived experiences shaping creativity and hindered the honed prideful skills of creative work.

\end{abstract}
% Problem: how do generative models interact wit users' values? 

% Approach: qualitative analysis

% scraped data from reddit
% value sensitive design 
% qualitative analysis
% reflexive thematic analysis

% Results:

% Conclusion:

\begin{CCSXML}
<ccs2012>
   <concept>
       <concept_id>10003120.10003121.10011748</concept_id>
       <concept_desc>Human-centered computing~Empirical studies in HCI</concept_desc>
       <concept_significance>500</concept_significance>
       </concept>
 </ccs2012>
\end{CCSXML}

\ccsdesc[500]{Human-centered computing~Empirical studies in HCI}

\keywords{Value sensitive design, generative AI, generative fill, values}

% \centering
%   \includegraphics[width=.9\textwidth]{Figures/chi-lbw-2023.pdf}
%   \caption{ }

% They are now less and less relying on instruction manuals or institutional support for learning technology or troubleshooting.


\maketitle

\input{intro}

\input{study}

\input{discussion}

\bibliographystyle{ACM-Reference-Format}
\bibliography{ianref}

\end{document}

%% file: intro.tex
%intro
\section{Introduction}

  % The creative industry is both concerned and enthusiastic regarding how generative AI will reshape creativity. How might these tools interact with workflow values or values of the creative arts? We use a value-sensitive design (VSD) framework on generative AI tools, particularly Photoshop’s Generative Fill (GF), supporting or hindering creative professionals' values. We obtained 2666 posts about "generative fill" from online forums where Photoshop is in their target users' work practices. Following data cleaning, we conducted a reflexive thematic analysis of 566 posts, focusing on usefulness, ease of use, and user values. GF users found the tool \textit{useful} in doing touch-ups, expanding images, and generating and tailoring prototypical images. GF helped their value of productivity by making work efficient. It created a value tension around creativity; GF helped reduce barriers to creativity but hindered the distinguishment of 'human' art. Finally, GF hindered lived experiences, like being in a rainforest in picture-perfect conditions, and the honed prideful skills of creative work, like difficult retouching, are disappearing in using generative AI tools.

In the last few years, generative models have had a tremendous impact on the world, including but not limited to creative professionals \cite{jiang2023ai, bender2021dangers, hacker2023regulating}. For example, AI has impacted healthcare \cite{zhang2023generative}, mental health support \cite{jo2023understanding}, software engineering \cite{weisz2022better}, and eduation \cite{baidoo2023education}, to name just a few.
Creative professionals are currently bombarded with a variety of new generative tools. At times there is a question of how creative users are adapting their work practices \cite{zamfirescu2023johnny, gmeiner2023exploring}. Research on scientifically advancing generative models \cite{chung2023promptpaint, feng2023promptmagician, brade2023promptify, lawton2023drawing} and their practical use as tools are occurring in parallel. In practice, AI tools in creative work such as  DALL-E, Stable Diffusion, Firefly, and Generative Fill have garnered both enthusiasm and concern among creative professionals \cite{DeCremer_MoriniBianzino_Falk_2023, Franzen_2023}. However, as the technological landscape evolves rapidly, several sociotechnical questions have emerged and remain unanswered. A major question is how these models and tools will align creative professionals' values in their workflows.
    
%3 para on lit review here
            AI-generated art is having a growing impact on the creative world, specifically appearing in the form of text-to-image (T2I) generation. T2I is the transformation of a text prompt input into a generated image. However such text prompts require prompt engineering, which is often a struggle. One study showed that individuals were challenged by systematic prompt design \cite{zamfirescu2023johnny}. Users experienced difficulties with generating prompts and evaluating the effectiveness of their prompts. Accordingly, several works have emerged on how to improve the prompt engineering process. To address struggles with T2I prompt engineering, the tool "RePrompt" was designed for refining T2I prompts \cite{wang2023reprompt}. RePrompt automates the process of transforming user-generated prompts by adding and removing parts of speech to increase the emotional expression of image generation. A similar work is the Promptify tool \cite{brade2023promptify}. This tool allows users to specify subject and style information and generate a more thorough image prompt. Other tools address the T2I method through further abstractions. PromptPaint offers an alternative approach borrowing from painting concepts \cite{chung2023promptpaint}. This tool introduces vectors between discrete semantics (interpolating between "cat" and "dog"), adding a directional semantic (shifting "dog" towards "fluffy"), as well as interventions during the generation process, all while utilizing analogies to art creation metaphors (mixing paints, layering paints, etc). Another T2I tool Reframer allows the user to prompt the AI for a drawing that is created with strokes, and then add or modify strokes alongside the AI \cite{lawton2023drawing}.

%  One of the biggest questions of generation creative artwork with AI is how to perform image generation, mainly, although not limited to "text-to-image" generation.

The experience of creative professionals with generative models is an important factor in the advent of generative models; as is the effect of generative models on society at large. One research group found that self-identified "creative professionals" were largely not worried about the existence of AI tools \cite{inie2023designing}. They found that a reason they \emph{were} excited was because of changes in productivity, a finding confirmed in our study. A focus on the relationship between writers and AI revealed the nuances of how writers relate to assistance from other writers versus how they relate to AI assistance \cite{gero2023social}. Similar to our work, these authors looked into the values of creative professionals. The authors noted that creative writers value intention, authenticity, and creativity and that particularly the values of authenticity and creativity impacted whether a writer would or would not consider using a computer for support. The existence of such tools also led to a variety of ethical concerns: the anthropomorphization of AI is problematic, in suggesting that the image generator is as much to credit for the result as a human creator; there is concern about the usage of generative AI to forge the style of artists; and, among other concerns, the uncertainty of the relationship between copyright law and training image generators \cite{jiang2023ai}. Beyond creative concerns, concerns about generative AI are prevalent across the field. For example, one set of authors worries that 1) the generative models being as large comes at a cost of money and carbon emissions, and 2) by training on large sets of data across the internet, these models will amplify a hegemonic world views, negatively impacting marginalized populations \cite{bender2021dangers}. The ethical landscape of generative AI is still clearly an important issue, which is why our work on generative AI offers another unique perspective, incorporating moral and ethical considerations as part of our process.

 % The experiences of creative professionals should be at the center of this conversation. Participatory design aims to center the user in the design \cite{inie2023designing}. Additionally, exploring how users choose when to consult generative AI as opposed to another human to augment their workflow is also useful \cite{gero2023social}. Examining the broader sociopolitical implications of the existence of AI, with a focus on artists provides an additional perspective \cite{jiang2023ai}.

% While we find all of these approaches beneficial, our approach is particularly rooted in the Value Sensitive Design (VSD) approach, introduced by Friedman \cite{friedman1996value}. The focus of this methodology is on stakeholders, who are those impacted by the technology, and what they consider important, with a particular focus on morals and ethics. For the focus of our study, we decided to consider as stakeholders the creators who actively produce creative work. To hone in further, we decided to look at Photoshop users, and a retrospective study on the introduction of the new Generative Fill (GF) AI tool.

In this paper, we discuss results from an empirical study of Photoshop's generative fill (GF), adopting a value sensitive design (VSD) approach \cite{friedman1996value}. VSD defines values as what is important to people in their lives, with an emphasis on ethics and morality, e.g., autonomy, ownership, and usability. We focus on currently active creative professionals, particularly Photoshop users, and examine how they interacted with GF---and through those interactions what values relationships are manifested. Specifically, we examined the following research question: 

\begin{displayquote}
\emph{How does the generative fill (GF) being useful for certain tasks, and making certain tasks easy to accomplish help or hinder users in embodying their values?}
\end{displayquote}

% Specifically, looking at how creative workers interacted with Generative Fill, we aimed to determine the value relationship between what tasks for which the technology was useful and easy to use, and how the tasks interacted with what the users considered important. We formalized our problem definition as follows:

Results from our study identified specific ways that GF was useful in users' work practices and was easy to use helped or hindered values of the creators. For example, GF users found the tool useful in doing touch-ups (e.g., removing blemishes), expanding images (e.g., changing dimensions or adding space for text), and generating and tailoring prototypical images (i.e., images based on an idea in the user's mind, e.g., stock photos). Consequently,  GF helped users with their value of productivity by making work practices efficient. However, a value tension was created around creativity: GF helped values of reducing the barriers to being more creative, but as a result hindered the value of distinguishability of `authentic' art created by a human without any AI assistance. Specifically, users languished about people gradually losing appreciation for creativity and creative work.

%% file: study.tex
\section{Methods}

% We found conversational data about generative fill via forum posts on Reddit and DPReview. We began by scraping data from these websites using the Google Search API and collected 2666 posts from threads that contained results for $<$Photoshop $+$ "Generative Fill"$>$. BERTopic was used to filter through the posts, collecting only those that the model was confident were about generative fill. Then, if a researcher reading a filtered post determined it was not about Generative Fill, the post was also discarded. A final set of 566 documents were determined relevant and were used for qualitative analysis. To analyze the documents, we performed iterative refined open coding and reflexive thematic analysis \cite{braun2012thematic}.

% Online conversations were collected to form our initial dataset. Forum posts were the focus of our data collection, 
% and the query \linebreak $<$ photoshop $+$ "Generative Fill" $>$ on the four domains. 

We collected data about GF use from online forums, including posts on Reddit r/photography, r/graphic\_design, and r/photoshop, as well as posts on the website DPReview. Data collection occurred on November 28th, 2023, using the Google search API. 2666 unique posts were obtained from threads that contained mentions of GF. We filtered out posts that were not relevant to our research question through a multi-step process. First, we used a randomized BERTopic model to identify posts that did not contain information relevant to generative fill. Next, the remaining posts were manually reviewed by the first author and eliminated if determined not relevant. When the relevance of a post was not immediately clear, additional review was conducted to examine contextual information, including finding the thread of ``replied to'' posts starting at a given post, and reviewing all posts by a particular user. A final corpus of 566 posts was then used for a qualitative analysis.

% The remaining posts were examined for qualitative analysis, but if a post was determined on observation to not be relevant, it was discarded. Although we examined text at the level of a single post in isolation, sometimes it was not immediately clear what the posts were discussing. To facilitate understanding, several tools were created to 1) find the chain of "replied to" posts starting at a given post, 2) show all the posts by a particular user and 3) search through all posts for a particular string. This was used strictly to provide context to unclear posts.
% We performed iterative refined open coding and reflexive thematic analysis \cite{braun2012thematic}.

% We elected to use reflexive thematic analysis~\cite{braun2021one} on the documents. The process consisted of a single researcher performing open coding on the documents, creating code groups, and through discussion with another researcher, iteratively refining the codes and code groups. The objective of this exercise was to produce codes that were relevant to easiness and difficulty of use, useful and non-useful aspects, and the value relationship with the posters. After a month of iterative coding, the result was 164 codes across 566 documents. Through further analysis including conversation among researchers, we were able to identify the themes in the relationship between the coded data and our research questions.

We used reflexive thematic analysis to analyze data from the 566 posts \cite{braun2021one, braun2012thematic}. \emph{ATLAS.ti} was used. A subset of the posts was open coded by the first author and then discussed and iterated upon in group data analysis sessions to define the scope of all subsequent analyses---different uses of GF, perceived ease of using GF, and user values embodying GF's use and usability. With that scope in mind, we focused later analysis (selective coding) on the ease and difficulty of using different features of GF, useful and non-useful aspects of GF, and user values manifested in their posts. After over two months of iterative coding, we identified four overarching themes that interweave to answer our research question.

% With that scope in mind, we focused later analysis (selective coding) on sources of tech support, preferences of tech support, challenges associated with different types of tech support, and perceptions about different types of tech support. After four months of iterative coding, we identified three overarching themes, 14 categories of codes, (Table \ref{tab:themes}), and 56 questions. 

% To that aim, we decided to focus on the following general areas: older users' identities, methods for sustaining and increasing proficiency, learning preferences, support preferences, and challenges associated with different learning preferences and methods.  Following six months of iterative coding, we finally arrived at 77 codes, which were then selectively coded into 11 categories (Table \ref{tab:codes}). 

\section{Results}
Following the thematic analysis of our data, we identified four overarching themes that addressed our research question about how different user values are embodied through GF use and usability. Some values were supported, some were hindered, while some competed against each other (value tension \cite{friedman1996value}).  Primarily, some users found that generative fill was useful in performing touch-ups to images (such as removing blemishes), expanding images, and generating prototypical images (borrowing the concept of idealized prototypes from psychology \cite{rosch1973natural}). This relationship helped and hindered users' values in a variety of ways. The inclusion of these useful features (touch-ups, etc.)  reinforced users' value of productivity by replacing complex repetitive tasks with simpler streamlined commands. Simultaneously, the useful aspects created a value tension within the spectrum of creativity -- both enabling users to do more with less and also adding the doubt of ``Was this made by an AI?''. The change of methods led to hindering both \textit{lived experiences} as well as a sense of accomplishment. Finally, generative fill inherently put designers in both a personal and interpersonal value tension where regulating content is either too loose or too restrictive. We conclude that the impact of this technology on creative workers is notable and that the resolution of value tensions may be found not only in the careful design of future tools but also in the broader response by society to sociotechnical progress.

%For example, it helped users values was that it enabled them to be more productive, where users were happy to find that they were able to get more work done with less resources. Creativity, on the other hand helped some users and hindered others: GF was helpful when it came to lowering the barrier to creative work, however it also hid the 

%We found that these capabilities helped the users to maintain their values of productivity, and created a value tension between generating prototypical images, and that it hindered their values of tangible work, job satisfaction, and verifiability. Additionally we noted that the value of users for respect for dignity and privacy (specifically related to the technology not being used for pornographic content, or "deepfakes") was in tension with other users' values for freedom of expression. 

% The themes we identified in our work are as follows: some users finding the technology easy for \emph{touchups, composition,} and \emph{generation} helped users in that it improved \emph{efficiency} and \emph{productivty}, and hindered users in that it decreased \emph{the need for photography, fulfillment,} and \emph{ability to trust information,} as well as created value tension around \emph{illicit material}.

\subsection{Generative fill is useful}
%\subsection{Touchups, Expanding Images, and Generating Prototypical Images}

The three tasks that we observed users describing as easy and useful, in a way that impacted their values, are performing touch-ups, generating prototypical content, and expanding images. \emph{Performing touch-ups} is defined as when a user makes a small change, such as removing a tattoo or removing powerlines to showcase the subject of the picture better. \emph{Generating prototypical content} is when a user has an image in mind and has to create a corresponding image to work with. \emph{Expanding images} is changing the dimensions of the image, providing space for marketing text on the image, or changing the background of images. Before discussing the value implication from these observations, we will first describe these tasks in more detail.

\subsubsection{Performing Touch-ups} Without generative fill, touch-ups are complicated, often involving several tools, and can take a significant amount of time \cite{Chen_2010}. With the tool, it is simplified. One user commented on the speed-up saying "30 min work in 30 seconds." We found users describing the difference with and without the tool:

\begin{quote}
"The time I spent doing selections, clone stamp, repainting, color match.... I can now do in seconds."
\end{quote}

Additionally, we found that users described how they could perform the task in a wide variety of cases. This speaks to the versatility of the method. For example, one user listed several ways in which they could perform touch-ups with the tool:

\begin{quote}
"I like to use gen fill to clean up movie and book covers, e.g. remove text labels, logos, watermarks."
\end{quote}
% \begin{quote}
%     "I use it for small touchups, usually without a prompt and for that it's fantastic."
% \end{quote}

% \begin{quote}
%     "Generative fill comes in very handy at times when you are doing touch-ups to a photo. Things such as getting rid of distractions. Especially small stuff."

\subsubsection{Generating Prototypical Content} Generating prototypical content is an ideal use case for text-to-image generative models. GF replaces the need for a creator to search through existing (stock) photographs until they find one that meets their needs. Surprisingly, the reaction of users to this particular task was mixed. For example, while some users approved of GF's ability to generate new images,

\begin{quote}
"With generative fill I can start with an empty frame and zero stock and end with something quite presentable."
\end{quote}

It was also the case that other users struggled with building up from a blank canvas or generating new images,

\begin{quote}
"If you're just trying to generate art from nothing, then I can see how you'd think it's useless. It's just not there yet."
\end{quote}

Workarounds to this problem included limiting the size of edits and using the tool as a "word brush". However, it is worth mentioning that there have been a number of research initiatives into redesigning the T2I prompting mechanism \cite{wang2023reprompt, brade2023promptify, chung2023promptpaint}, indicating that perhaps the current selection and text prompt mechanism could be further evaluated for usability.

%Finally, users reported that they could generate prototypical content (often replacing the need for  using image generation, which they found easier to use. One user describes the new found ability with this technology:

% And another user recounts the amount of time it saved them compared to searching through stock photograph:

\subsubsection{Expanding Images} The ability to extend images received a wealth of positive reactions, being referred to as "jaw-dropping", "godly", and a "game-changer". The use of the technology simply involved selecting the area to expand the image and prompting (or leaving the prompt blank). A user describes the process as follows:
\begin{quote}
"Say I'm developing a vertical poster, and the image I want to use across the full height is horizontal. You can frame the image how you want it and generate the gaps in art."
\end{quote}

The user feedback on expanding images was consistent. Wth touch-ups or prototypical images, there were the occasional concerns that users raised such as, "It's been useful for cleanup work but even then, it hasn't done anything I can't do myself with honestly more predictable results." On the other hand, image extension appears to lack the same divisiveness, except for concerns that exist for all usage of the technology, such as limitations on the resolution of the images that can be generated.

% Additionally, several users reported that they were frequently using the tool to expand images, allowing them to easily deal with starting images that don't have enough room to work with, or when they need the dimensions changed for a specific orientation.

% \begin{quote}
% "My life was struggling with content aware fill to expand on the cropped photography clients kept handing over like it was ready for use. Leaving no space for text, logos, USP's, what have you.  Now, with generative fill, it's SO SIMPLE to expand a photo as much as I want. Really a game-changer for our team."
% \end{quote}

%In the following subsections, we discuss several of these value relations. Some of the relationships are very obvious (these tools being useful means a user will be more productive) while some are more nuanced (these tools replace the need for other types of work).

\subsection{Generative fill helps with productivity}

We define \emph{productivity} as "being able to accomplish tasks with less time or effort." We observed this to be a key value in users of generative fill, an observation which has been repeatedly noted elsewhere \cite{Rao_unk, inie2023designing}. ." Users who valued productivity were concerned with effectively getting the desired output. One user discusses how they saw this from GF as the replacement of complex and time-consuming operations with a quicker and easier workflow. They said,

\begin{quote}
"It speeds my workflow tremendously eliminating a ton of clone stamp/healing brush manipulation that is, frankly, a waste of my time."
\end{quote}

Additionally, users often associate productivity with materialistic gains, such as an ability to have an increased earning potential or decreased work hours. One user describes their experience of both gains, resulting from their increase in productivity.

\begin{quote}
"AI has literally turned my 10 hour day into a 2 hour day. And my work has improved to the point where I'm making more money in 2 hours than I was in 10"
\end{quote}

\subsection{Creativity: helped or hindered?}

Making the technology easier to learn and work with, or \emph{reducing barriers to entry}, is another observed value of stakeholders in the technology. In fact, we found that there were a number of users who explicitly highlighted this proliferation as a value. Two users spoke to this saying,
\begin{quote}
    "It’s a creative medium, people will create what they want to and this just makes it easier for many to manipulate images in lots of ways."
\end{quote}
and
\begin{quote}
    "This tool just cuts down the time spent on it dramatically, as well as opening up the ability to do it to people who don't have years experience mastering every little trick." 
\end{quote}

However, in contrast with the benefits to less experienced users, there were also concerns that \emph{AI art is indistinguishable from human work}. We observed users who highlighted that such increases in creative capacity leads to an indistinguishability, cheapening the value of their work,

\begin{quote}
"From a hobbyist standpoint everyone’s “art” will look the same. If everything from rough idea to finished work is done by a program what’s the fucking point? This is stupid."
\end{quote}

One user spoke to the duality of the problem,

\begin{quote}
"It both terrifies and impresses me equally. Photography and static art in general as we know it is going to change in one instant more than ever before."
\end{quote}

In VSD this is what is known as a value tension, where multiple values exist that are in opposition but potentially could be made to coexist. \cite{friedman1996value}. Since these two values are in tension, resolving the tension would mean finding a solution that addresses both the concerns around authenticity of human art while still keeping the creative opportunities afforded by GF.

% It also leads to the question of whether these results will be generalizable to other generative models, where a strike between enablement and cheapening not only in creative image generation but in other applications may also exist.
% \subsection{Productivity and Creativity}

% We found that some users were able to complete tasks faster, allowing them to create more content, or spend more time focused on work they found meaningful. For example, one user said:

% \begin{quote}"Instead of spending 3 hours on menial, time-consuming things, you can now spend that time fine tuning your design and making it even better in less time."
% \end{quote}

% Additionally, we found that some users claimed the technology, by replacing the need for explicit photoshop skill, made creativity more accessible to more people. One such remark was: 

% \begin{quote}"It’s a creative medium, people will create what they want to and this just makes it easier for many to manipulate images in lots of ways."
% \end{quote}

% Lived Body and Enjoying The Craft
\subsection{The Changing craft: a feeling of loss}

Values related to what can emerge from the existence of the tool tell only part of the story. The aspects of life that are lost with the emergence of new technology are also a key part of the user experience.

The experience of the \emph{"Lived Body" is potentially lost} if technologies like GF replace the need for "in the world" work. "Lived Body" comes from the work of Merleau-Ponty \cite{merleau2013phenomenology}. In his phenomenological philosophy, he considers the experiences of the body such as the quality of the air, the smells and sounds, and the overall feeling of being in a place, as essential components of how we perceive the world. Being present in an environment with a camera is one instance of this. Some users claim that the inspiration for being in a location provides its own value beyond that of the quality of the image. One user accounts how this is important, should it be possible,

\begin{quote}
"[I]f I were to be doing a photoshoot and wanted it to be in the rainforest and could afford it, I'd want to go to the freaking rainforest. Not just for how it looks, but the feeling and inspiration that comes with shooting on location."
\end{quote}

Simultaneously, GF also has the potential to make obsolete old modes of work, which would lead to a \emph{loss of pride in craft}. Practitioners with Generative Fill  claim that they enjoy the work (a user says "I enjoy retouching" and another says that the new technique "sounds really boring"). Loss of pride in craft is observed as the notion that people took pride in their method of work and enjoyed the method. In comparison, one user languishes over the new method of work,   

\begin{quote}
"Quite frankly I don’t feel the same sense or accomplishment upon completion either because it’s just so easy."
\end{quote}

% Some users talked about the enjoyment and experiential benefit of working in the tangible world, something they worried would be replaced with working behind a desk. A demonstration of this value is:

% \begin{quote}
% "[I]f I were to be doing a photoshoot and wanted it to be in the rainforest and could afford it, I'd want to go to the freaking rainforest. Not just for how it looks, but the feeling and inspiration that comes with shooting on location."
% \end{quote}

% Other users talked about a lack of satisfaction from the methods of generative fill, compared to the satisfaction of working with previous photoshop tools. It was stated, \begin{quote}"Typing a prompt into a generator until I get the least wonky looking end product sounds really boring."
% \end{quote}

% Another concern that some users had with the technology was that with the barrier to entry lower, it would be easier to make fake images for the user of misinformation and propaganda, and it would be harder to verify if an image was real or fake. One such claim was,

% \begin{quote}
% "[T]o the untrained eye, AI goes unnoticed. not sure how it's gonna go, just the AI is going to make fake news stuff alot more fake."
% \end{quote}

% \subsection{Respect for Dignity and Privacy and Freedom of Expression}

\subsection{Respect for dignity and privacy and freedom of expression}

Finally, we found that there is a value tension that exists around what GF is capable of, and how it is regulated. Particularly, the question is how the ethical decision should be made if the creation of content is allowed or prohibited. Adobe addressed this issue directly in the technology with the creation of "guidelines" that would block the use of GF if it determined the user was attempting to make illicit content. 

The two values clearly conflict with each other concerning explicit content. The respectful use of technology, not creating unsettling content or directly manipulating someone's person without their consent, maintains individuals' \emph{sense of dignity and privacy}. And simultaneously there is the belief that an individual should be \emph{free to create content} without interference if they are using it for legitimate means.

Adobe's guideline restrictions prevent unethical use of the technology \emph{by} preventing the creation of the type of images that could be considered explicit. As an example specifically of how this led to values coming into tension, we look at two users who took the issue from two different perspectives. The first spoke to the risk of the feature being used for fake pornographic images of people, which should be censored,
% A final concern that users had with the technology is that it would be easier for users to create illicit content. For example, one user worried:

\begin{quote}
"I think it would be a very bad idea to have a feature in Photoshop that allows you to easily make high-quality fake nude images of people."
\end{quote}

While this is a valid instance for censorship, sometimes censorship is used when the use case is legitimate. For example, how do you choose to censor nudity in the case of classical art which often works with nude models? One user comments on their frustration with getting censored repeatedly as they are trying to go about their work,

\begin{quote}
"I know a lot of us work with nudity and it's extremely annoying that we are getting an insane amount of guidelines violation messages when working on those pictures."
\end{quote}

 We found that a large number of users were bothered by guideline violations (the current solution to the problem of illicit content), so it would seem reasonable to expect that a more nuanced solution is still necessary.

%% file: discussion.tex
\section{Future Work}
The value of creative arts to society is fundamental \cite{milbrandt2010understanding, stuckey2010connection}. Creativity is a uniquely human quality
\cite{balter2009origin}. Today, however, generative AI applications can produce new content in the form of text, images, audio, and video, or a combination of those. While some creators are experimenting with these tools to augment their creative work, some are alarmed at how they can impact their lives and livelihoods, e.g., by creating unfair competition \cite{Walton_2024a}. Some have even speculated a future ``techlash'' against algorithmically generated content, where people begin to value authentic creativity more and be willing to pay a premium for ``human-made''\cite{DeCremer_MoriniBianzino_Falk_2023}.  This new value, the ability to create, recognize, and appreciate art created by a human, without any AI assistance, is then shaped not due to how the technology affects the creators or consumers, per se, but rather the position the technology places them in. Thus, it is important to understand how the work practices of creative professionals are evolving with the evolution of generative AI tools and which of their values are being helped, hindered, or are in tension. 

In this paper, we reported results from a preliminary study on how users' values are embodied in ways that they find Photoshop's Generative Fill (GF) meaningful to their workflow. As a future step, it is important to understand how individual differences in creative professionals influence their value perceptions when using generative models in their work practices. For example, how age, experience, and expertise would influence how people use generative tools in creative work? And, as a result, how would workforce training, skill-building, hiring, and mentoring practices change? Would prompt engineering be a required skill in graphic design in the future? How would the lived experiences of people play a role in how they use generative models in creative work practices? On the other hand, it will also be interesting to study how these tools can help people who are not in the creative profession express their creativity better. And when doing so, how can these tools uphold their values, like becoming an expert with colors or mastering portrait photography?

% \begin{itemize}
%     \item Productivity was greatly increased; users were able to significantly speed up their work, and utilize the change in workflow to meet professional needs
%     \item Barriers to entry were decreased; the difficulty in creating digital art shrunk, enabling artists with less talent to make more compelling images
%     \item Verifiability of artistic work decreased; artists who engage in Photoshop use without AI are now at risk of their work being indistinguishable from work created by an AI
%     \item Experience of "Lived Body" was lost; photographers generally prefer to perform their work on sight where they can fully experience the world of their work. With the advent of generative AI, the need to do so has decreased, and the opportunity to be in those places has decreased with it
%     \item The enjoyment of the craft was lost; some users found satisfaction in the act of sorting through tools and putting in the effort to create an image. Some users viewed this as a meaningful process that they regret being lost
%     \item Respect for dignity and privacy is at risk; the inherent capabilities of generative models pose the threat of generating unethical images of people without their consent, and
%     \item Freedom of expression can potentially be enhanced; where the desire to create images freely, including explicit content if the situation calls for it, is possible with generative models, even if in its current form it is heavily moderated.
% \end{itemize}